\documentclass[conference]{IEEEtran}
\IEEEoverridecommandlockouts
\usepackage{cite}
\usepackage{amssymb,amsfonts}
\usepackage[fleqn]{amsmath}
\usepackage{algorithmic}
\usepackage{graphicx}
\usepackage{textcomp}
\usepackage{xcolor}
\usepackage{lscape}
\usepackage{multirow}
\usepackage[para,online,flushleft]{threeparttable}
\usepackage{algorithm,algorithmic}
\usepackage{boldline}
\usepackage{caption}
\usepackage{subcaption}
\usepackage{mwe}
\usepackage{wrapfig}
\usepackage{hyperref}
\usepackage{nameref}

\def\BibTeX{{\rm B\kern-.05em{\sc i\kern-.025em b}\kern-.08em
    T\kern-.1667em\lower.7ex\hbox{E}\kern-.125emX}}
\begin{document}

\title{Billion-files File Systems (BfFS): A Comparison}

\author{\IEEEauthorblockN{Sohail Shaikh}
\IEEEauthorblockA{\textit{Department of Computer Science} \\
\textit{George Mason University, Fairfax, Virginia}\\
mshaikh5@gmu.edu}
}

\maketitle

\begin{abstract}
As the volume of data being produced is increasing at an exponential rate that needs to be processed quickly, it is reasonable that the data needs to be available very close to the compute devices to reduce transfer latency. Due to this need, local filesystems are getting close attention to understand their inner workings, performance, and more importantly their limitations. This study analyzes few popular Linux filesystems: EXT4, XFS, BtrFS, ZFS, and F2FS by creating, storing, and then reading back one billion files from the local filesystem. The study also captured and analyzed read/write throughput, storage blocks usage, disk space utilization and overheads, and other metrics useful for system designers and integrators. Furthermore, the study explored other side effects such as filesystem performance degradation during and after these large numbers of files and folders are created.
\end{abstract}

\begin{IEEEkeywords}
Linux, filesystem, EXT4, XFS, BtrFS, F2FS, ZFS, performance, statistics
\end{IEEEkeywords}

\setlength{\textfloatsep}{5pt plus 1.0pt minus 2.0pt}
\setlength{\floatsep}{8.0pt plus 2.0pt minus 2.0pt}
\setlength{\intextsep}{10.0pt plus 2.0pt minus 2.0pt}


\section{Introduction}
The local filesystem, where the operating system and applications reside, forms the core of many systems. As local hard drives scale up, the number of objects stored in each node grows. This study is to investigate the hypothesis that there are likely some limitations and performance impacts of storing and accessing large numbers of files (e.g., a billion files) of varying sizes on popular local filesystems such as EXT4, XFS, BtrFS, F2FS, ZFS, etc. The study intends to evaluate different local filesystems available for the Linux operating system, their pros and cons, their ability to store a billion files and directories, and their suitability for the purpose under specific workloads on different types of storage media such as SSD and HDD.

Linux EXT4, the 3rd revision of the EXT filesystem has been around since the mid-2000 and has performed very well as a general-purpose and stable local filesystem for popular Linux distributions. However, as the data volume has been increasing rapidly with large amounts of data being placed closer to the compute devices to improve the performance of data-intensive computations and avoid data transfers over the network, the importance of local filesystem performance and their limitation has increased rapidly. The core idea of this study is to answer two basic questions supported by analysis: a) Can these filesystems handle one billion or more files and folders? and b) does the performance deteriorates as the number of files and folders are added? If unable to store a billion files, analyze and determine if these limits can be overcome.

\textbf{EXT4} is currently the default filesystem for most popular Linux distributions. EXT2 and earlier filesystems were prone to catastrophic corruption when the system loses power during a write operation.  EXT3 solved this problem by journaling, a 2-phase writing process of metadata and the data. EXT3 was limited to a 2TB file size and a maximum 16TB filesystem size. EXT4 increased this limitation to 16TB file size and 1EB filesystem size. There are other improvements in EXT4 such as different types of block allocations, extents, and an unlimited number of subdirectories using HTree indices\cite{b16}, the checksum for journals, nanosecond timestamp, and online defragmentation\cite{b7}.

\textbf{XFS} is a high-performance 64-bit journaling filesystem and was ported to the Linux kernel in 2001. XFS excels in the execution of parallel input/output (I/O) operations due to its design\cite{b4}. XFS design enables extreme scalability of I/O threads, filesystem bandwidth, size of files, and the filesystem itself when spanning multiple physical storage devices. XFS ensures the consistency of data by employing metadata journaling and supporting write barriers. Space allocation is performed via extents with data structures stored in B+ trees, improving the overall performance of the file system, especially when handling large files.

\textbf{BtrFS} is a modern modified Btree-based Copy-on-Write (CoW) filesystem where files are not replaced but a new copy with updates are created. It is aimed at implementing advanced features while also focusing on fault tolerance, repair, and easy administration. Its main features and benefits are snapshots, RAID, and self-healing. 16EB file size, dynamic inode allocation, SSD awareness, automatic defragmentation, and scrubbing\cite{b6}.

\textbf{ZFS} is another filesystem developed by Sun Microsystem and acquired by Oracle, combines the roles of volume manager and a file system intended for critical systems where data loss is not acceptable. ZFS is intended for large infrastructures where large number of storage devices are deployed. ZFS requires appropriate configuration to support these storage devices.

\textbf{F2FS} (Flash-Friendly File System) is a filesystem specifically intended for NAND-based flash memory devices equipped with Flash Translation Layer (FTL). It is supported by Linux kernel 3.8 and above and supports compression as well as encryption natively. F2FS was designed around the idea of a log-structured filesystem approach, which is adapted to newer forms of storage and overcame issues such as the snowball effect of wandering trees and high cleaning overhead. F2FS has a weak \textit{fsck} that can lead to data loss in case of a sudden power loss\cite{b8}.

\section{Methods}
To compare target filesystems, they needed to be exercised thoroughly and compared against a baseline. The baseline is established by creating 10 million files of different sizes that are \textit{normally} distributed between 1KB and 10KB, and stored evenly among 100 root folders. To establish a baseline, the first step is to set up the environment for these experiments. Due to several factors including the introduction of unpredictable internet and ISP-induced latencies, and response time from the cloud components, it was decided to avoid the cloud altogether and rely on the on-prem server hardware for experiments. The specification of the on-prem HP Z820 workstation is: 2x Xeon E5-2670 2.6GHz CPUs with 16 cores/32 threads total, 256GB RAM, dual NICs connected to two independent subnets, one 1TB SSD for Linux Ubuntu 22.04, a 2TB Samsung 870 EVO SATA SSD\cite{b12} with ~550MB/sec read/write performance for applications, and a 14TB Seagate IronWolf™ HDD with 256MB cache and a 210MB/s sustained transfer rate\cite{b11}. In addition, C/C++ development tools, Java OpenJDK 11, Visual Studio, and other tools were installed such as filebench, fio, vmstat, iotop, dstat, atop, visualjvm, and considered using bcc\cite{b17}. The goal is to capture IOPS, read and write performance, and I/O latency of reading and writing files. To exercise these filesystems with files and folders reads and writes, an application was developed. The operating system was not fine-tuned in any way deliberately to make this study reproducible and useful for others.

\subsection{Application}
After compiling \textit{filebench} and making several attempts to test and failed at around 10 million files mark due to crashes, it became evident that it is time to write some purpose-built tools. Initially, Java environment is used for its flexibility and consistent performance to develop the benchmarking and evaluation application\cite{b2}. To ensure that application overheads are consistent regardless of the number of files to be created or read, simple programming constructs were used. In addition, to eliminate any performance degradation due to introduction of a JVM, the application is also (re)developed in C language using the lowest level Linux file system calls\cite{b19}. This paper uses performance metrics from C application. The high-level pseudo code for folders and files creation and verification are shown in Algorithm \ref{algorithm1}.

\begin{algorithm}
  \caption{Folders/Files create/write/read algorithm}
  \label{algorithm1}
  \begin{algorithmic}[1]
    \STATE $Files \leftarrow$ 10M or 100M or 1B files
    \STATE $Folders \leftarrow$ 100
    \STATE $Subfolders \leftarrow$ computed from $Files$ 
    \FOR{$i=1$ \TO $Folders$}
        \STATE Create folder
        \FOR{$j=1$ \TO $Subfolders$}
          \STATE Create subfolder
          \FOR{$k=1$ \TO $100$K}
            \STATE Create file of normal random data size with crc-32
          \ENDFOR
        \ENDFOR
        \STATE $Statistics \leftarrow$ collect folder/file statistics        
    \ENDFOR
    \STATE $Statistics \leftarrow$ collect storage statistics
    \FOR{$i=1$ \TO $Folders$}
        \STATE Search folder
        \FOR{$j=1$ \TO $Subfolders$}
          \STATE Search subfolder
          \FOR{$k=1$ \TO $100$K}
            \STATE Read file and verify content using crc-32
          \ENDFOR
        \ENDFOR
        \STATE $Statistics \leftarrow$ collect folder/file statistics
    \ENDFOR
    \STATE Process collected statistics for analysis
  \end{algorithmic}
\end{algorithm}

The application consists of two parts: a \textit{Creator} and a \textit{Reader}, routines that create folders, subfolders and files and read folders, subfolders and files, respectively. To accommodate 1 billion files with a planned file size range of 1KB to 10KB, a 14TB HDD was used. The rationale behind the file size range is that a typical disk block size is 4KB so a file may use anywhere from one and up to three 4KB blocks~\cite{b3}. During file creation, each file is added with random binary data and is appended with an 8-byte CRC-32 checksum generated on-the-fly so that the \textit{Reader} can verify the file’s content during reading.

The Creator routine also captures file sizes and individual write throughput distribution during the runs. At the end of the run, it summarizes captured metrics which are then fed into an Excel spreadsheet for further analysis. The Reader performs the reads, iterating thru the folders and files created by the Creator earlier. The Reader reads each file's content, extracts the last 8 bytes of the content, the CRC-32 checksum, computes the CRC-32 checksum of the remaining content, and compares the two to ensure file's integrity. The application counts the number of failed checksum and reports the file count discrepancy for the folder as well as for the entire run at the end. Following is the files/folders schedule that is used for testing these file systems: $Files = Folders * Subfolders * 100$K.

\begin{table}[h]
\caption{Files/Folders Read/Write Schedule}
\label{tab:read_write_schedule}
\centering
\begingroup
\setlength{\tabcolsep}{6pt} 
\renewcommand{\arraystretch}{1.5} 
\begin{tabular}{rrrr}
\hline
\textbf{Files} & \textbf{Folders} & \textbf{Subfolders/Folder} & \textbf{Files/Subfolder} \\ 
\hline
\small{10M} & \small{100} & \small{1} & \small{100,000} \\
\small{100M} & \small{100} & \small{10} & \small{100,000} \\
 \small{1B} & \small{100} & \small{100}& \small{100,000} \\
\hline
\end{tabular}
\endgroup
\end{table}

In addition to reading and writing files and folders, the application captures metrics for \textit{macrobenchmarking} and \textit{microbenchmarking}. Macrobenchmarking captures read/write performance metrics for the individual folders and files contained in them as well as the entire run at the end whereas microbenchmarking focuses on individual file read/write performance and distribution (min, ave., max). It is important to note that captured timings are in microseconds.

\subsection{Linux Filesystem}

Focusing only on Linux operating system for this study, Linux file system is designed around several abstract layers built on top of each other for ease of interfacing and to isolate the applications from unnecessary complexities. Applications access block devices, such as hard disks through system calls (e.g., \textit{open()}, \textit{read()}, \textit{write()}) to the Virtual File System (VFS). VFS is the software layer in the kernel that provides the filesystem interface to the user space programs[5] and interfaces with one or more installed file systems such as EXT4, XFS, and others. Between the hardware and file system, there exist two more layers: the page cache and the block layer.

Linux page cache layer is the main disk cache and improves file system performance by caching data to and from the disk. The kernel first looks in the cache to see if the data is available and only accesses the disk if the data is not found in the cache. Once the data is read from the disk, the page cache is updated. In the case of a write operation, data is checked if it is already in the cache, if not a new entry is added but the data is not written immediately to the disk to allow accumulating more data before committing to the disk, reducing the I/O overheads. 

\begin{figure}[htbp]
\centerline{\includegraphics[scale=0.35]{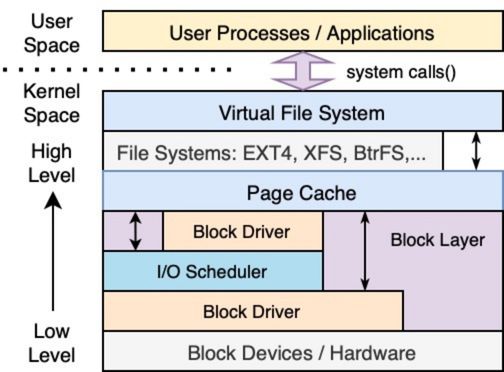}}
\caption{Linux VFS Architecture}
\label{fig:VFS}
\end{figure}

This delayed write to the disk is handled by the I/O scheduler in the block layer (which improved write speed). Block drivers hide the complexity of the underlying storage hardware e.g., hard disks (HDD), solid-state disks (SSD), optical disks, etc. In the context of this paper, there are a few other important aspects of the filesystem implemented in Linux VFS that are useful to understand for this study and explained in the following sections, e.g., directory entry cache, inode, and file object. Read/Write latency for the file system is described as follows where each \textit{layer} contributes to some amount of latency: \textit{Latency\textsubscript{read/write} = Latency\textsubscript{application} + Latency\textsubscript{VFS} + Latency\textsubscript{filesystem} + Latency\textsubscript{driver} + Latency\textsubscript{diskIO}}. 


Application latency is largely dependent on the programming language and constructs to access the filesystem, and any high-level abstractions used to make the filesystem transparent to the programmer. Linux Virtual File System (VFS) latency is related to abstracting the filesystem’s internal working from the high-level system utilities as well as the applications so that different filesystems look the same to the applications\cite{b10}. 

However, the latency imposed by the application is assumed to be constant or with minimum variations between runs. VFS also imposes a negligible overhead as compared to the amount of time spent in other layers. Since the application, block driver, and block device layers are constant therefore the only layer left is the filesystem which is the layer this study tries to capture and analyze. The study has analyzed the portion of run time that is consumed by the application, other software layers, and by the hard disk I/O.

\subsubsection{Directory Entry Cache}
File-related system calls such as \textit{open()} and \textit{create()} require \textit{path} as an argument. This argument is used by the VFS to search for the directory entry in a lookup cache, called Directory Entry Cache (\textit{dcache}) and it provides a view into the entire filespace. Due to the constraint of limited physical memory, entries (\textit{dentries}) are sometimes created on the fly using inode information.

\subsubsection{Inode Cache and Inode}
Inode cache (\textit{icache}) go together with \textit{dcache}, in a master-slave configuration. If there is a \textit{dentry}, there is an entry in \textit{icache}. The reason \textit{icache} exists in the first place is that the \textit{inode} gets updated frequently while the file is open but the inode is saved on the disk unlike \textit{dcache}. An inode or index node exists for one per object in the filesystem for all object types (files, directories, etc.) It is a structure and holds not only the inode number but also the file size, owner’s user id, and group id, access and creation times, etc.

\begin{figure}[htbp]
\centerline{\includegraphics[scale=0.3]{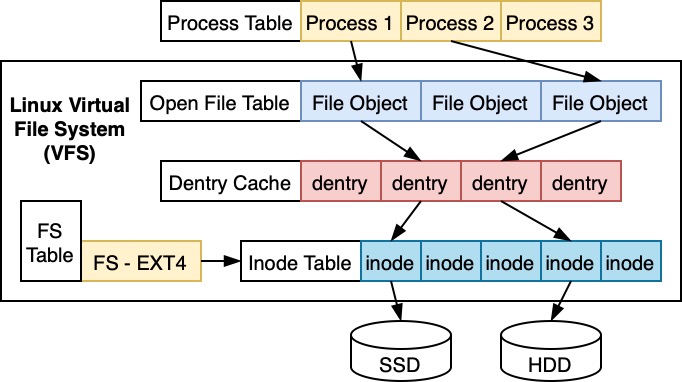}}
\caption{EXT4 Internal Structures}
\label{fig:inodes}
\end{figure}

\subsubsection{File Object}
For each opened file in a Linux system, a file object exists and contains information such as its path, \textit{dentry} entry, file operations, flags, permission, mode, and other metadata[10].  Each file is assigned a file descriptor (FD) which points to the file object and is used for file operations. Linux reserves the first three FDs for standard input, output, and error and imposes a configurable 1024 FDs per-process limit.

\subsection{Discovery Approach}
To determine the read/write behavior of the target filesystems namely: EXT4 (baseline), XFS, BtrFS, ZFS and F2FS, several tools were explored but nothing satisfied the goals of this study. To overcome this roadblock, an application is developed in Java and later in C language which creates files and folders and captures performance metrics for further analysis. The targeted metrics are described below for writes and reads.

\subsubsection{Baseline Metrics - EXT4\textsubscript{10M} filesystem}
As the first step, a baseline needs to be established for comparison before conducting any further analysis. EXT4, which comes out of the box with most Linux distributions, is used as the baseline filesystem that allows further analysis and comparison of targeted filesystems to support one billion files. Application program Creator and Reader ran on the 14TB HDD and produced the results listed in Table \ref{tab:baseline_disk_before_and_after}.

\begin{table}[h]
\caption{EXT4\textsubscript{10M} Baseline Disk Metrics}
\label{tab:baseline_disk_before_and_after}
\centering
\begingroup
\setlength{\tabcolsep}{6pt} 
\renewcommand{\arraystretch}{1.5} 
\begin{tabular}{lrrr}
\hline
\textbf{Parameter} & \textbf{Before} & \textbf{Used} & \textbf{After} \\
\hline
\small{Inodes} & \small{1200005235} & \small{10000200} & \small{1190005035} \\ 
\small{Blocks} & \small{3171666459} & \small{19270745} & \small{3152395714} \\
\small{Disk size} & \small{1.2991$\times 10^{13}$} & \small{78932971520} & \small{1.2912$\times 10^{13}$} \\
\hline
\end{tabular}
\endgroup
\end{table}



\subsubsection{File Writes}
The application provides the capability to create folders and then create files between the sizes of 1K and 10K following a normal distribution curve, using Box-Muller transform\cite{b18}, that produced a consistent file sizes with the median falling around 5500 bytes with std. dev. of 1024 bytes. Reason to chose this range of file sizes is to ensure that one billion files fit in the 14TB hard disk as well at least 66\% of the file sizes are over 4KB block size occupying at least two blocks (except in the case of ZFS which uses 128KB blocks). 


\begin{wrapfigure}{r}{0.25\textwidth}
  \begin{center}
    \includegraphics[width=0.25\textwidth]{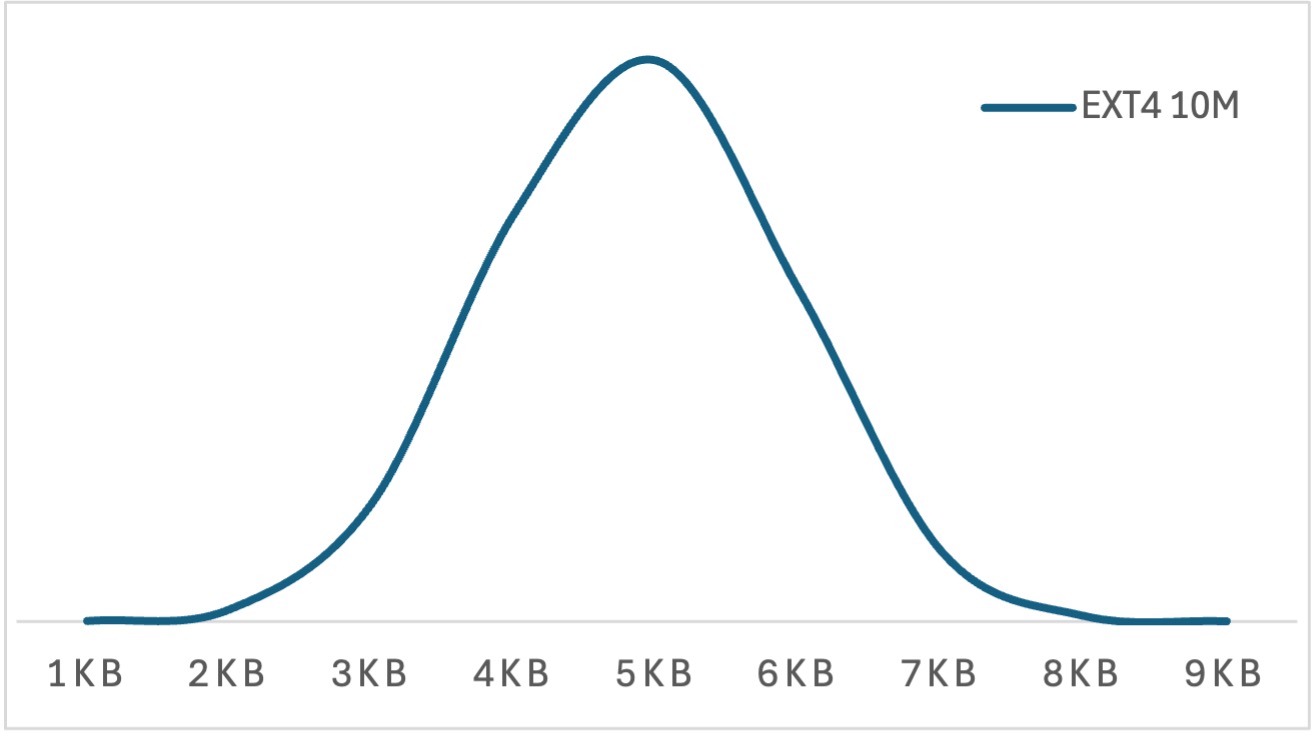}
  \end{center}
  \caption{File Size Distribution}
  \label{fig:file-size-distribution}
\end{wrapfigure}

The file size distribution pattern is shown in the Figure \ref{fig:file-size-distribution}. The tool also captured additional information to assess the performance. Individual files write time is aggregated for each folder and used for cumulative write time for the entire run. It is shown in Table \ref{file-write-performance-metrics} as the Total Write Time (TWT) in $\mu$sec. Captured individual file write time is also used to determine the min, max and average file write times. Total bytes written on the disk are captured to determine the disk utilization and overheads of adding the files. The tool captures an exact number of bytes per file and per folder which are added to get the cumulative size written to the disk for each run. After each run, inode count, blocks used, and increase in the disk space utilization are captured to determine the overheads.

\subsubsection{File Reads}
In order to understand the differences between write performance and read performance, we have to consider that during the write operation the cache is used therefore the system calls return quicker than in the case of reads since the application reads all the bytes in the memory, computes CRC-32 check, and compare with what is provided in the file before proceeding to the next file. File systems provide different read performance considering number of files generated earlier. For this comparison, the paper consider the average write and average read statistics. The captured average write performance is between 10\~15 $\mu$sec whereas read performances varies between 5 to 200 $\mu$sec. Except in the case of ZFS, the difference between write speed and read speed of the same files was not too great, e.g., on average EXT4\textsubscript{100M} write is 24 $\mu$sec whereas read is 62 $\mu$sec which is 3 times slower. However, for ZFS\textsubscript{100M} write is 16 $\mu$sec but the average reads are 201 $\mu$sec, a 12 times slower performance.

\subsubsection{Disk Utilization Overheads}
Disk utilization overhead is another aspect of the file creation process where the disk space used is greater than the bytes added, which is calculated by \textit{((Disk Space Used - Bytes Added) / Disk Space Used) * 100}, and has been observed to be anywhere from 30\% to 60\% for different filesystems. However, except in one case (i.e., XFS) number of files and folders did not affect this overhead because of its handling of folders and files metadata dynamically.

Though total folder creation time is relatively small, the tool captures these metrics to ensure it is not high enough to skew the analysis, especially for larger runs e.g. 10M files and above. Per folder write throughput which is determined by individual file write time divided by the number of files per folder. It is used to analyze performance degradation as the number of folders and files are added to the filesystem. 

\begin{figure*}  
\begin{subfigure}{0.24\textwidth}
\includegraphics[width=\linewidth]{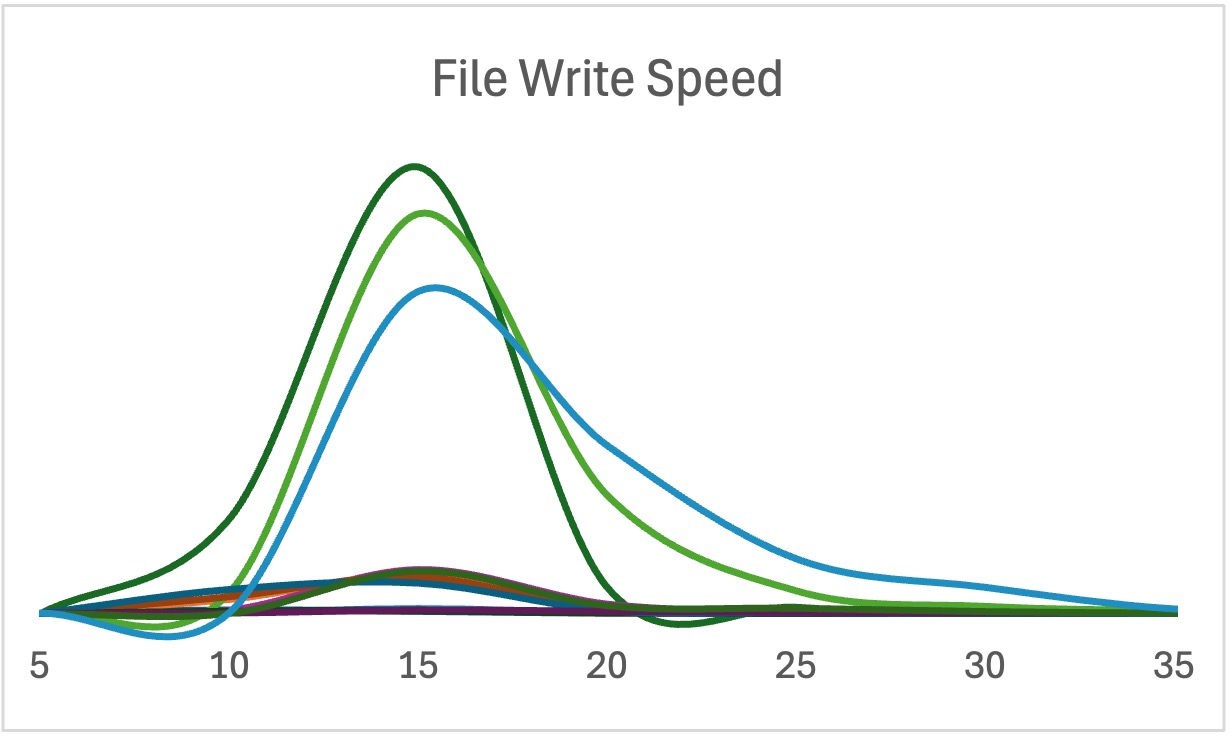}
\caption{All Files}
\end{subfigure}
\begin{subfigure}{0.24\textwidth}
\includegraphics[width=\linewidth]{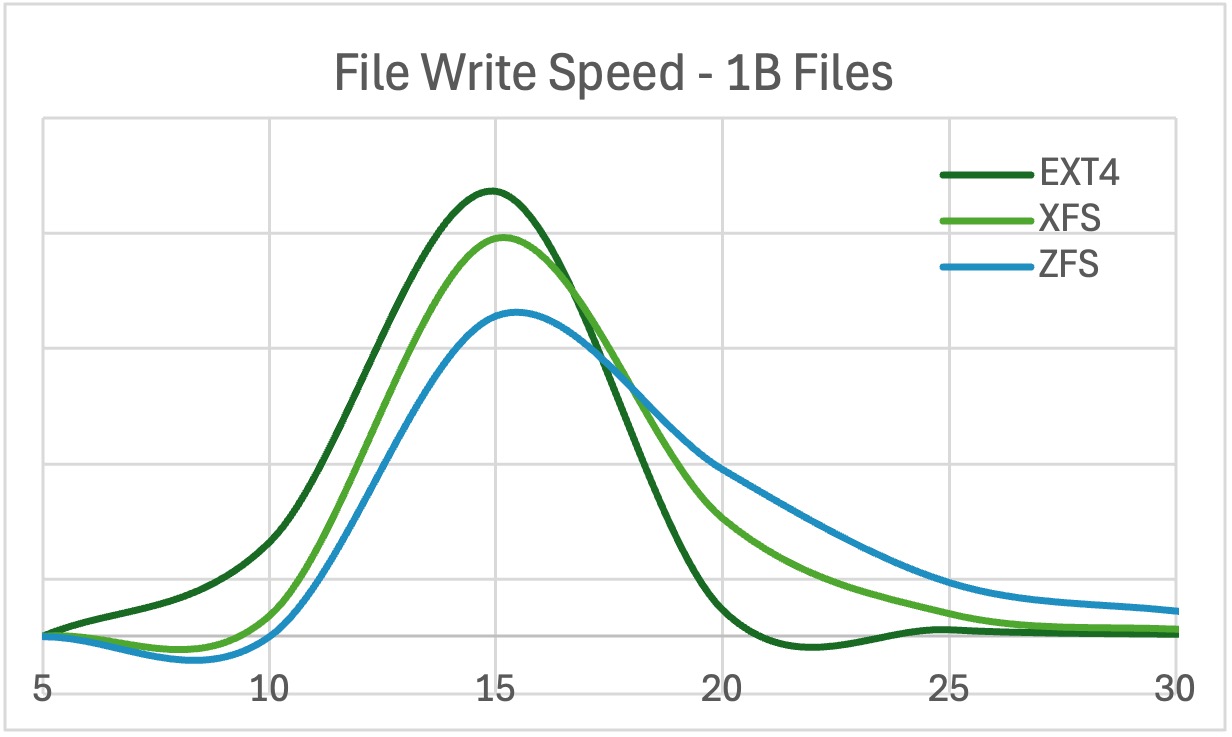}
\caption{1 Billion Files}
\end{subfigure}
\begin{subfigure}{0.24\textwidth}
\includegraphics[width=\linewidth]{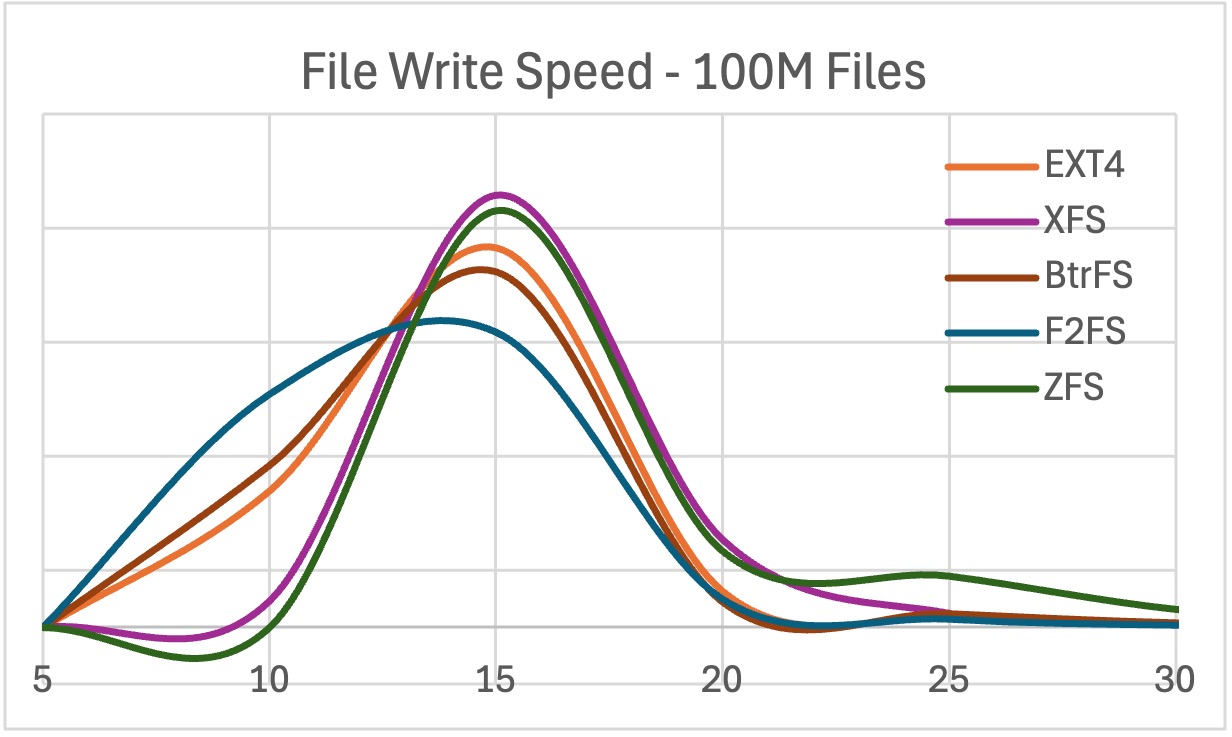}
\caption{100 Million Files}
\end{subfigure}
\begin{subfigure}{0.24\textwidth}
\includegraphics[width=\linewidth]{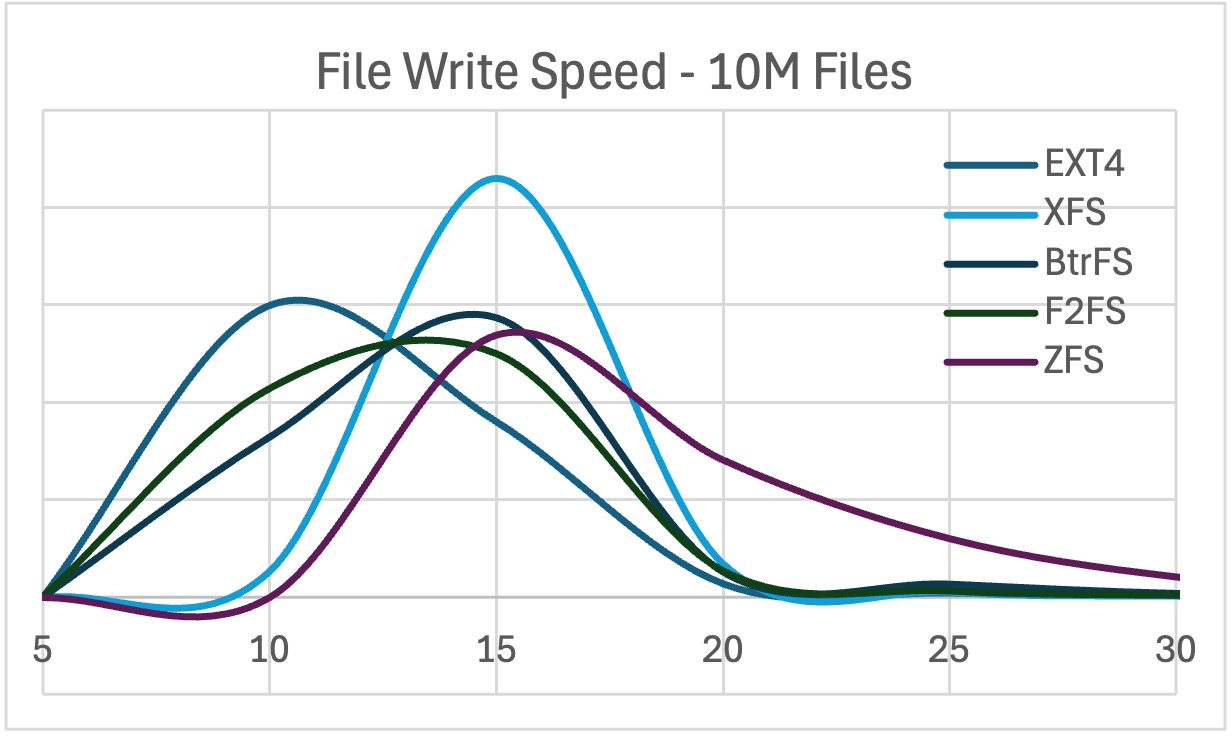}
\caption{10 Million Files}
\end{subfigure}
\caption{File Write Speed Distribution (in $\mu$sec)} 

\end{figure*}
\begin{figure*}  
\begin{subfigure}{0.24\textwidth}
\includegraphics[width=\linewidth]{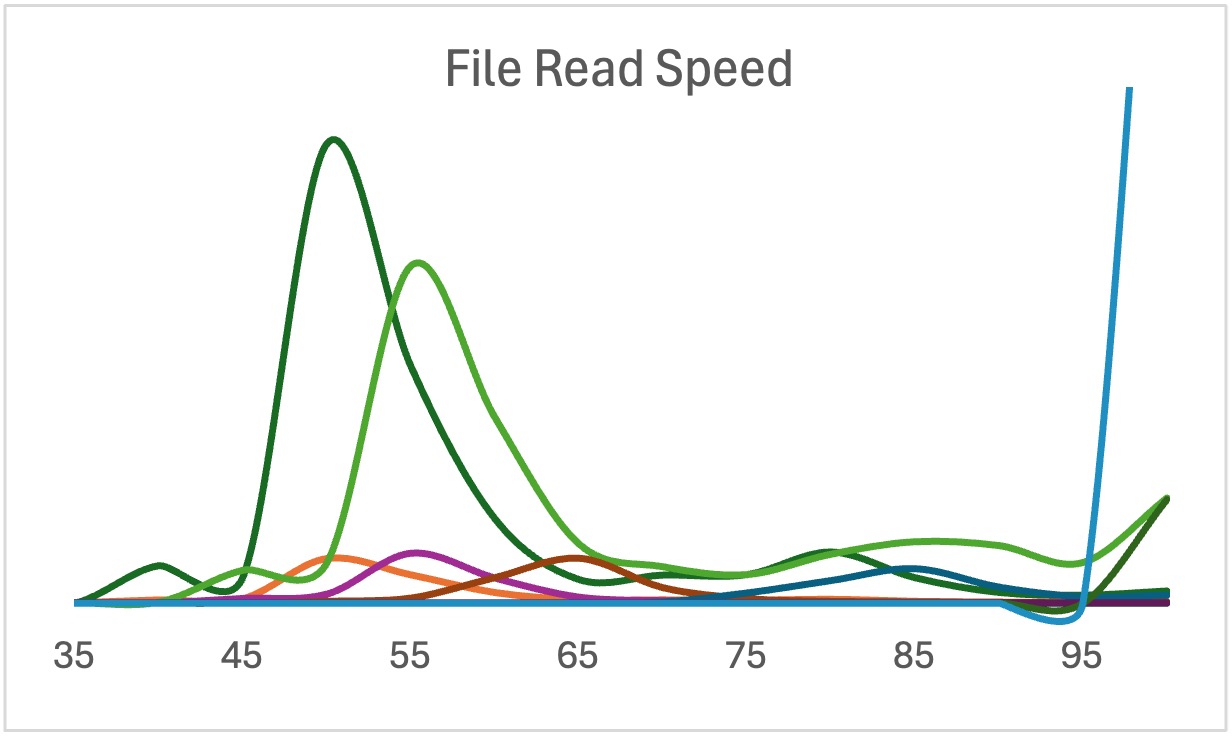}
\caption{All Files}
\end{subfigure}
\begin{subfigure}{0.24\textwidth}
\includegraphics[width=\linewidth]{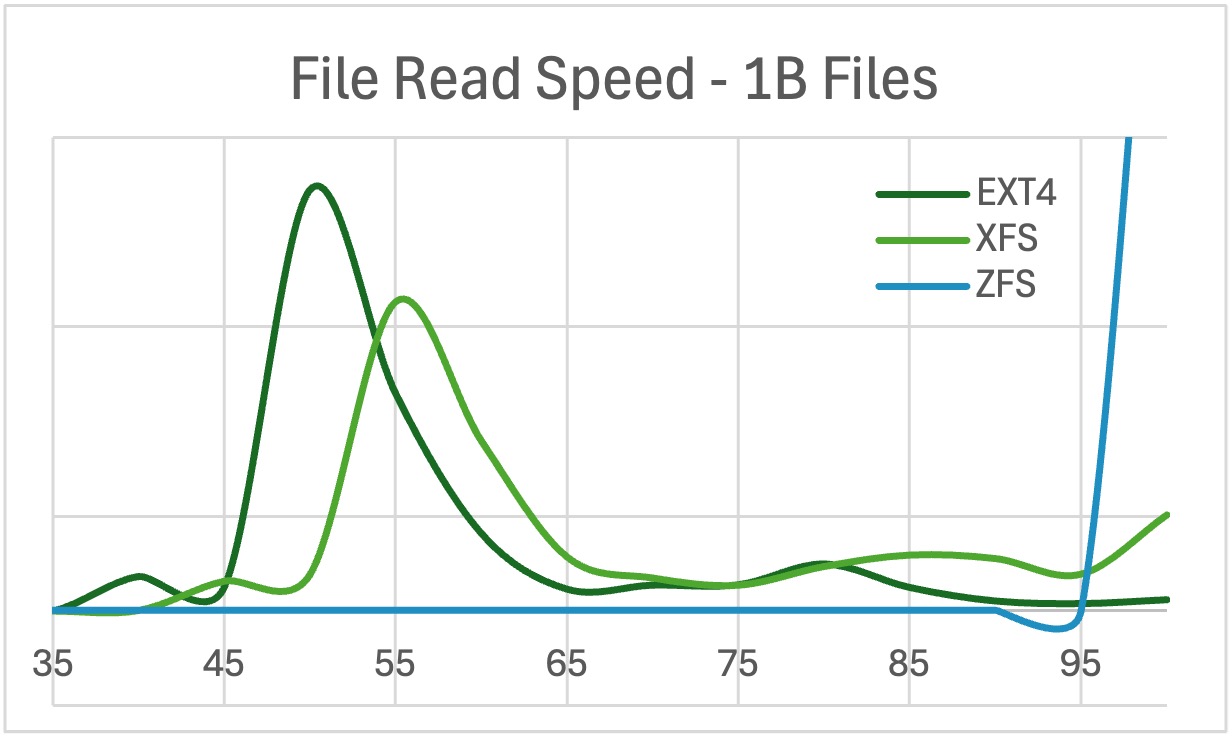}
\caption{1 Billion Files}
\end{subfigure}
\begin{subfigure}{0.24\textwidth}
\includegraphics[width=\linewidth]{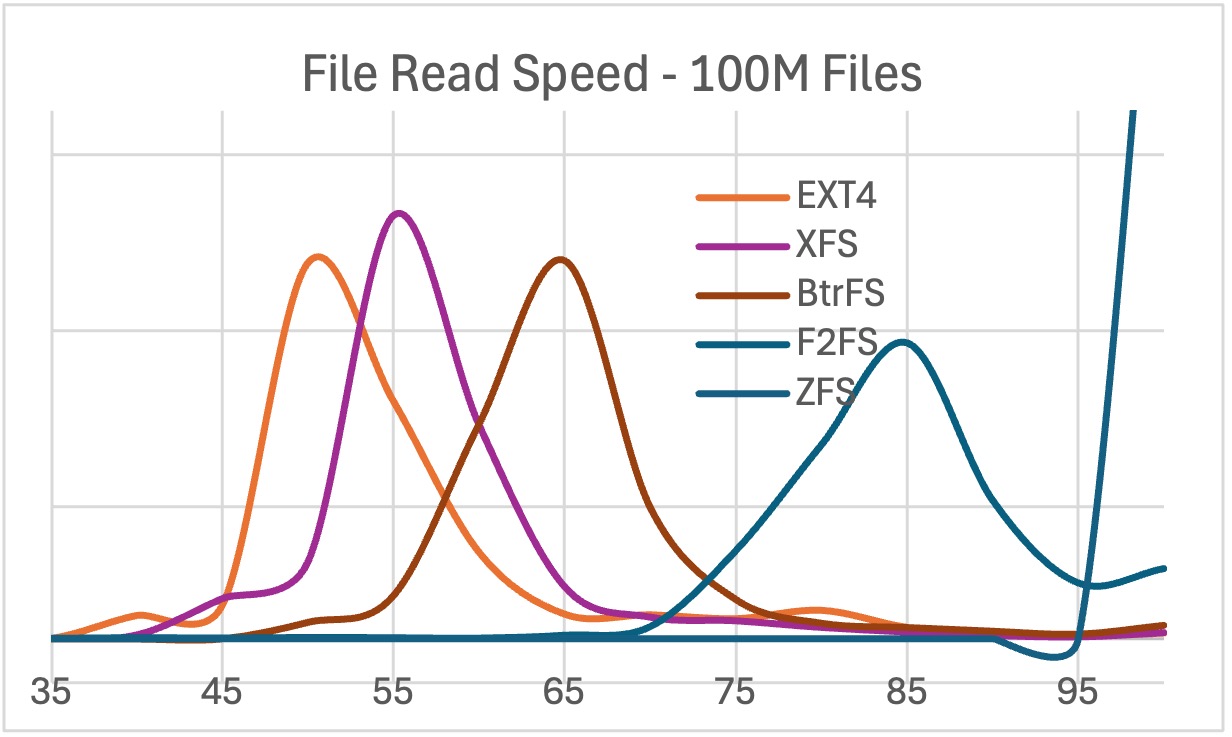}
\caption{100 Million Files}
\end{subfigure}
\begin{subfigure}{0.24\textwidth}
\includegraphics[width=\linewidth]{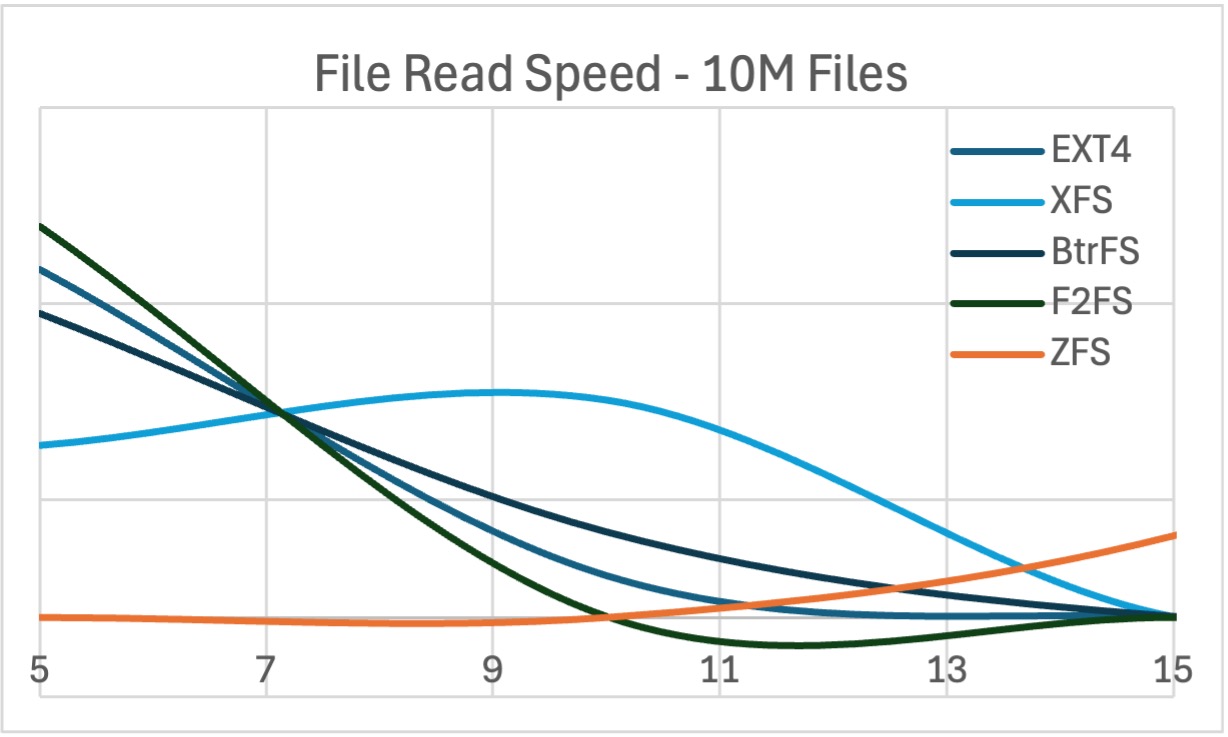}
\caption{10 Million Files}
\end{subfigure}
\caption{File Read Speed Distribution (in $\mu$sec)} 
\end{figure*}

\begin{figure*}  
\begin{subfigure}{0.24\textwidth}
\includegraphics[width=\linewidth]{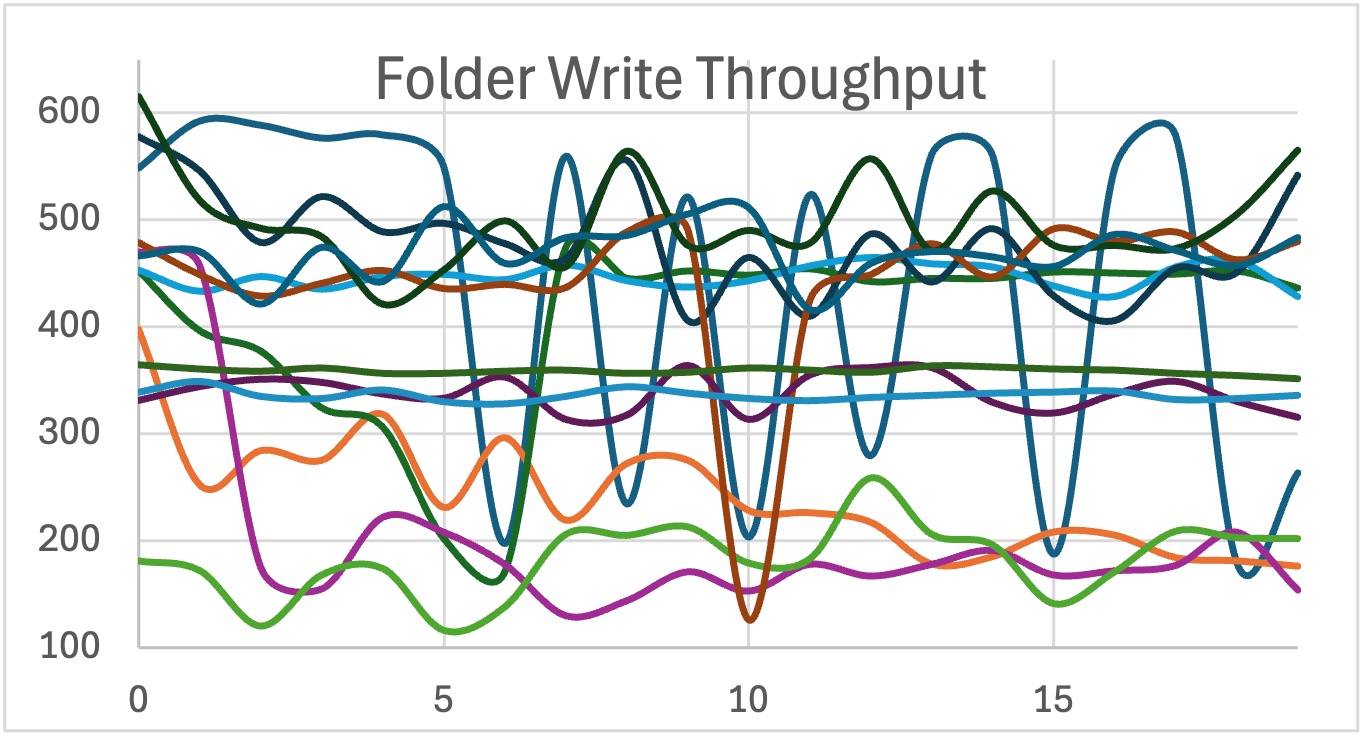}
\caption{All Files}
\label{folder-write-throughputs-all-files}
\end{subfigure}
\hfill 
\begin{subfigure}{0.24\textwidth}
\includegraphics[width=\linewidth]{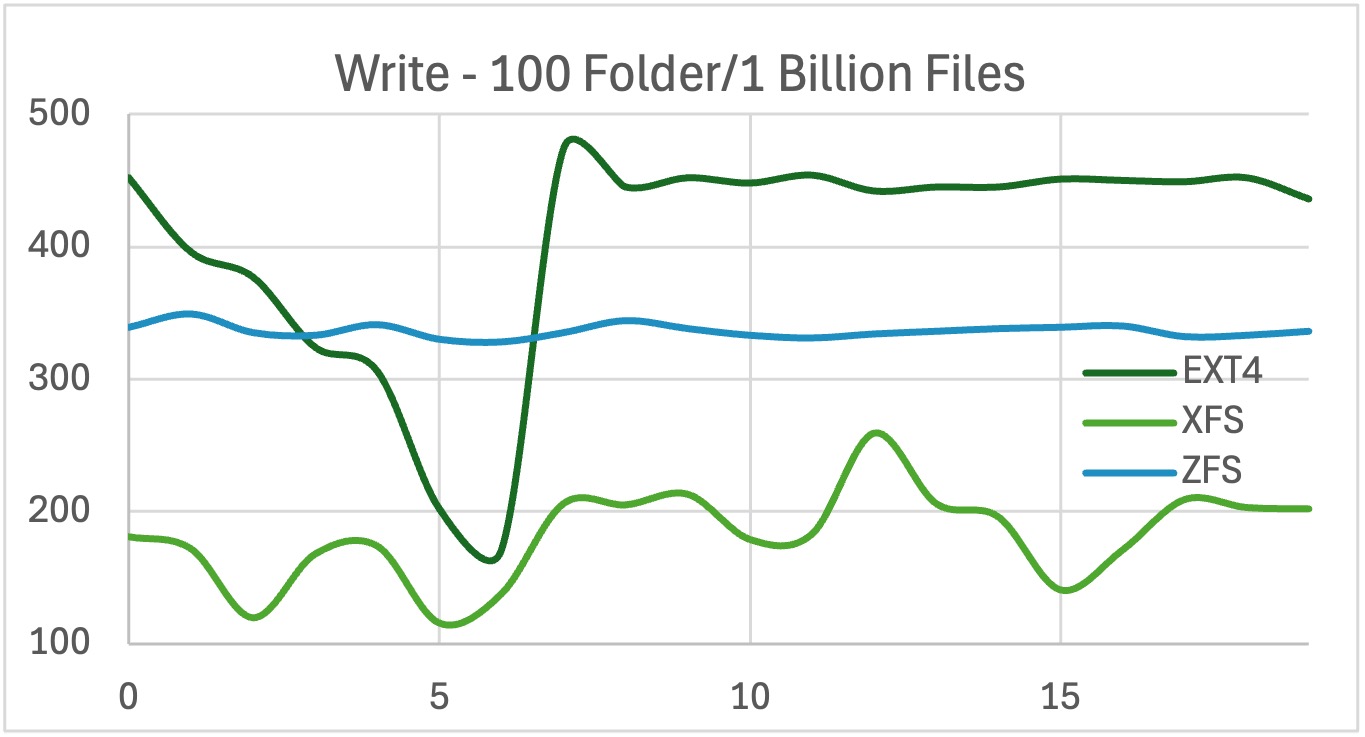}
\caption{1 Billion Files}
\label{folder-write-throughputs-1B-files}
\end{subfigure}
\hfill 
\begin{subfigure}{0.24\textwidth}
\includegraphics[width=\linewidth]{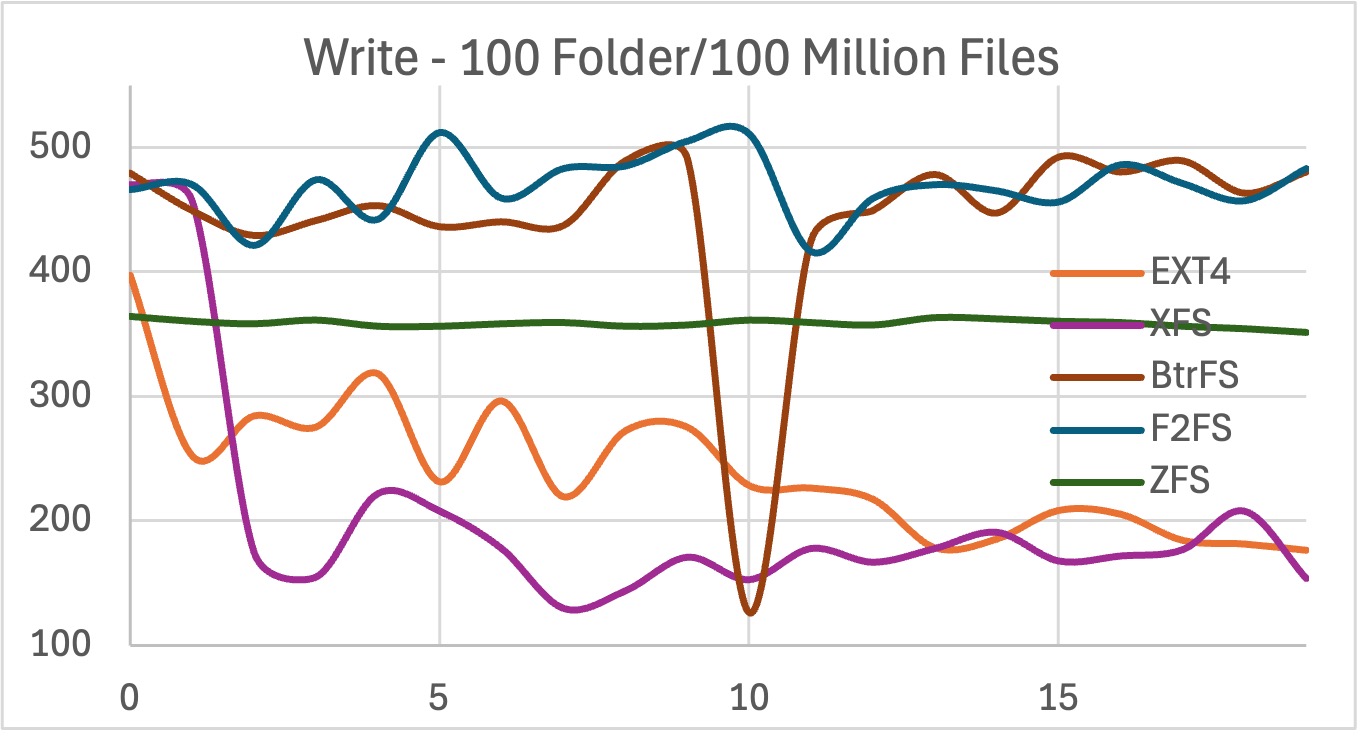}
\caption{100 Million Files}
\label{folder-write-throughputs-100M-files}
\end{subfigure}
\hfill 
\begin{subfigure}{0.24\textwidth}
\includegraphics[width=\linewidth]{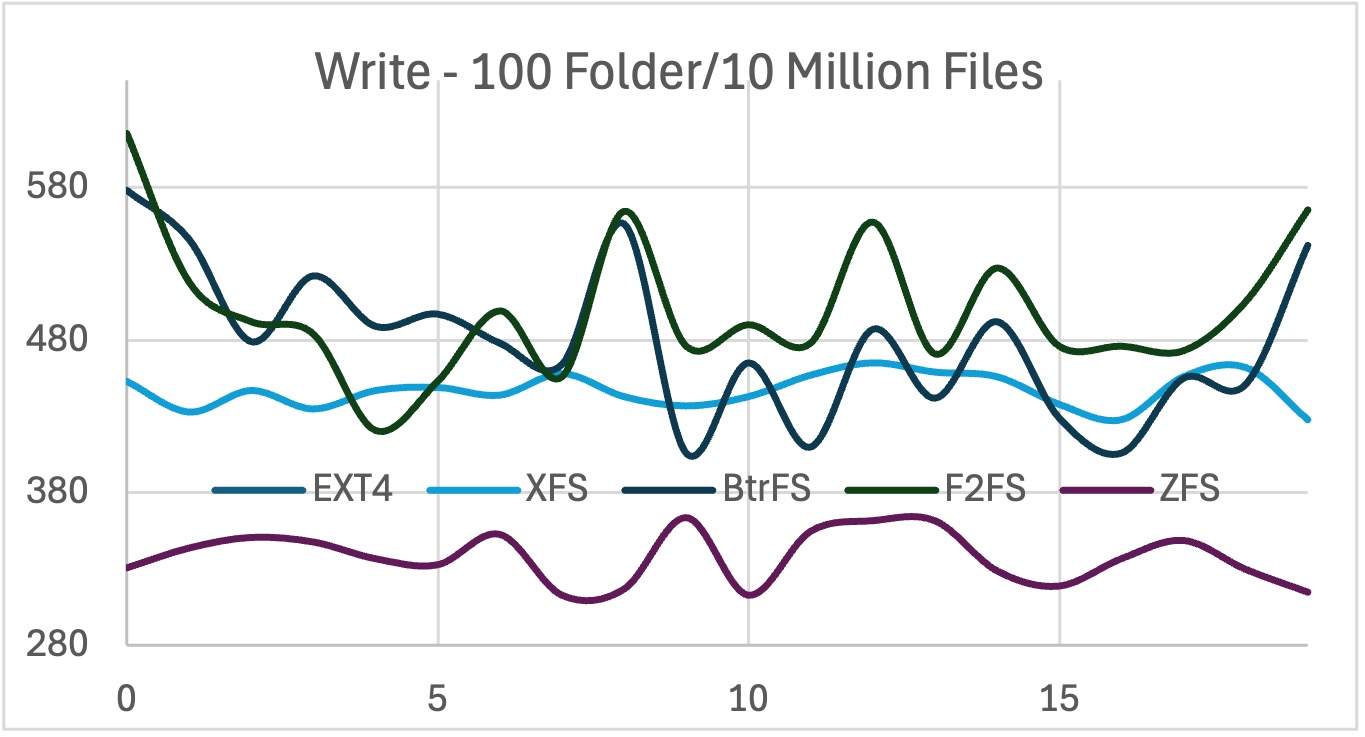}
\caption{10 Million Files}
\label{folder-write-throughputs-10M-files}
\end{subfigure}
\caption{Folder Write Throughput Distributed Over Time ($\mu$sec vs 20 buckets)} 
\end{figure*}
\begin{figure*}  
\begin{subfigure}{0.24\textwidth}
\includegraphics[width=\linewidth]{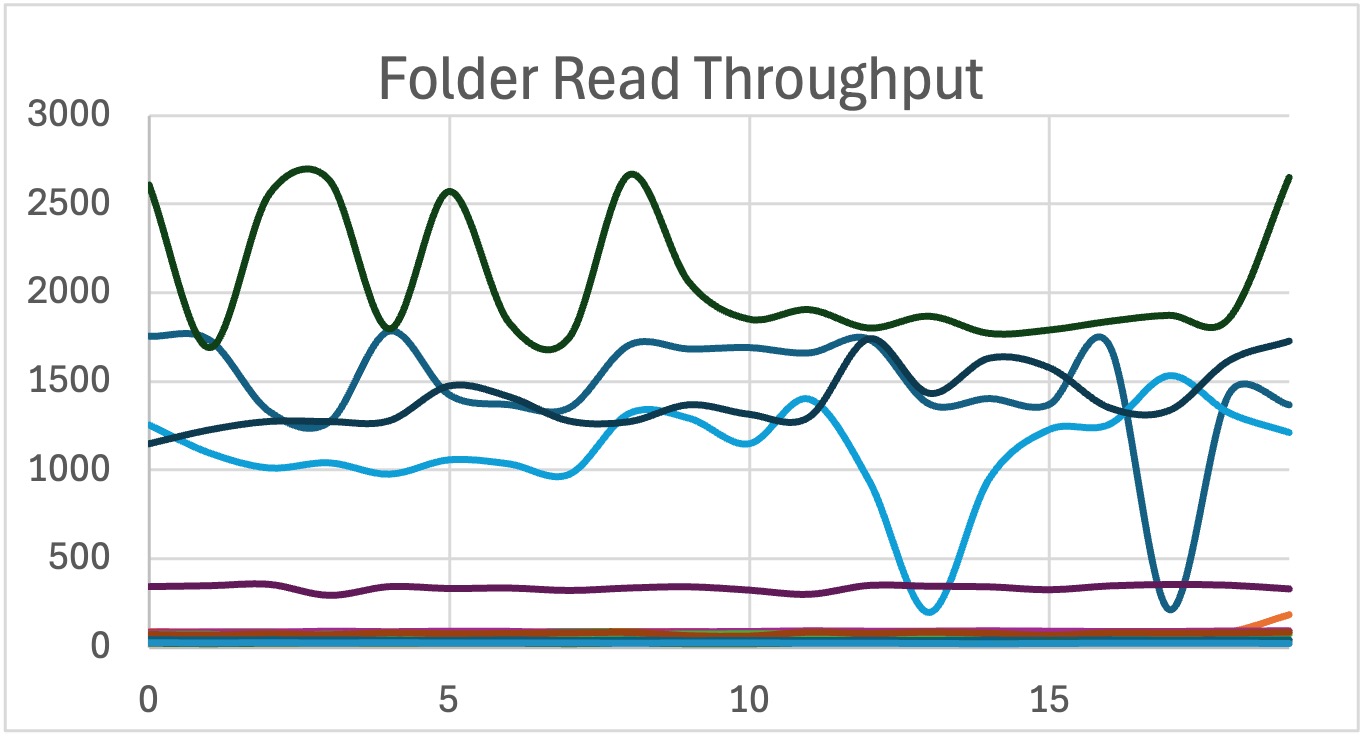}
\caption{All Files}
\label{folder-read-throughputs-all-files}
\end{subfigure}
\hfill 
\begin{subfigure}{0.24\textwidth}
\includegraphics[width=\linewidth]{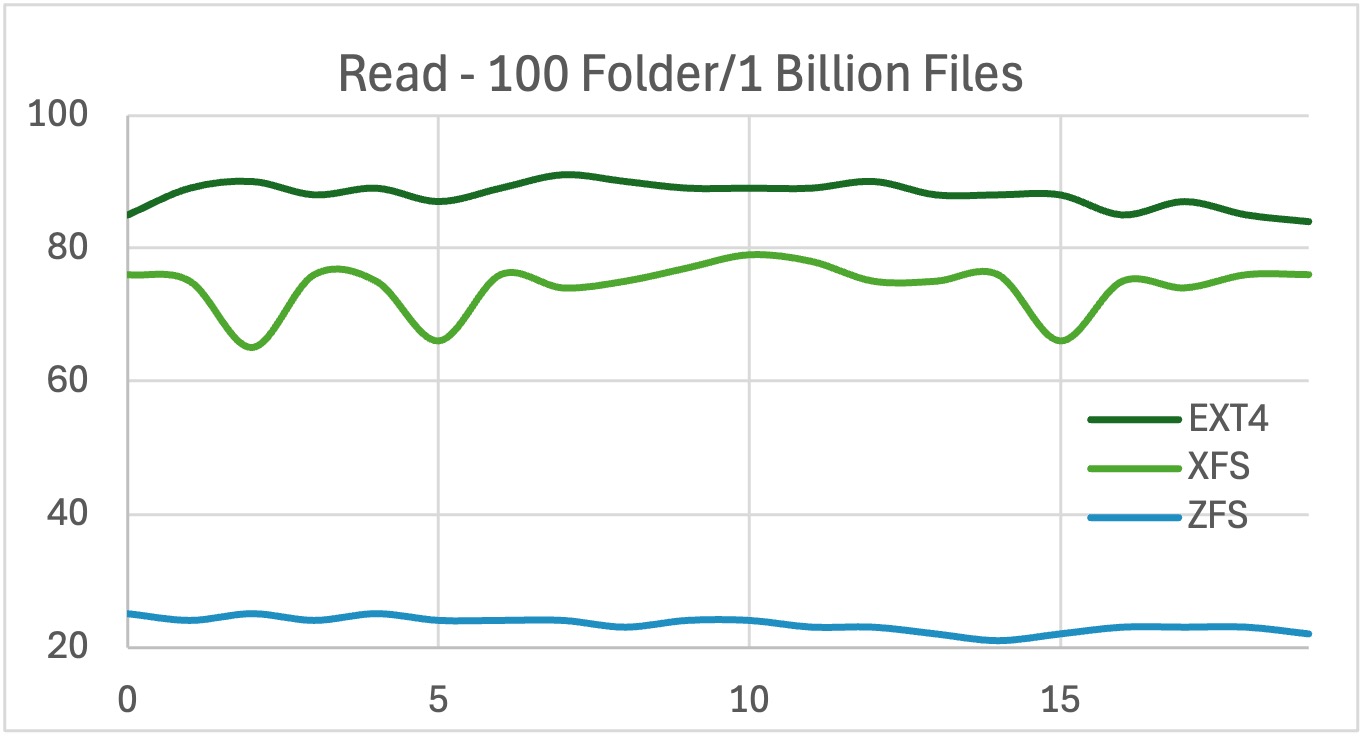}
\caption{1 Billion Files}
\label{folder-read-throughputs-1B-files}
\end{subfigure}
\hfill 
\begin{subfigure}{0.24\textwidth}
\includegraphics[width=\linewidth]{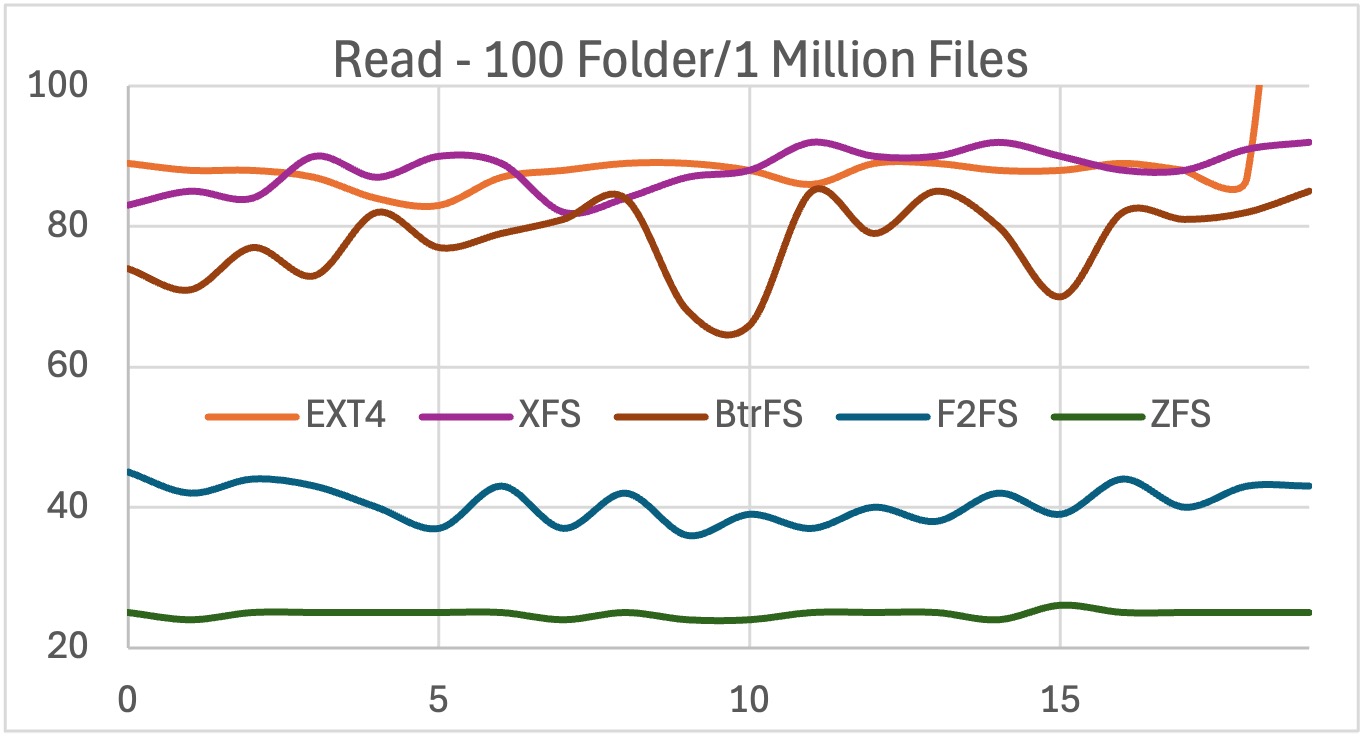}
\caption{100 Million Files}
\label{folder-read-throughputs-100M-files}
\end{subfigure}
\hfill 
\begin{subfigure}{0.24\textwidth}
\includegraphics[width=\linewidth]{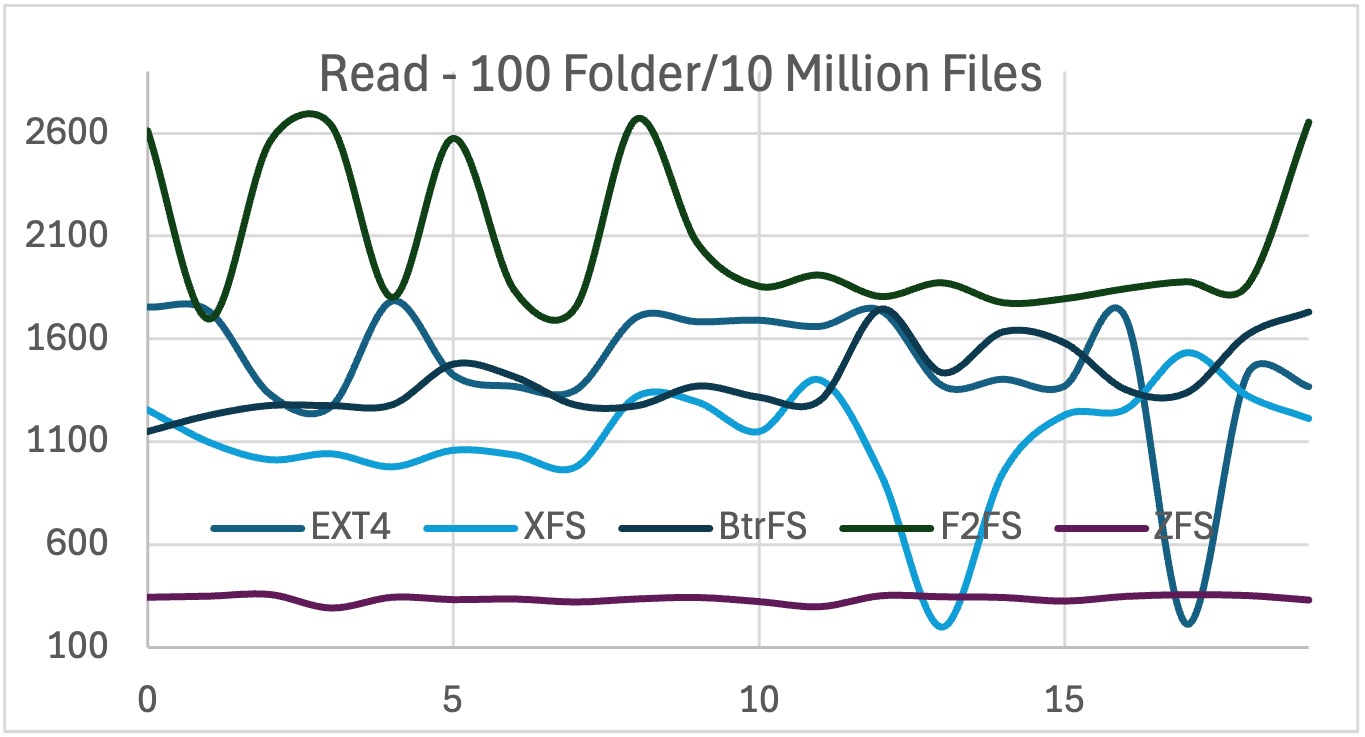}
\caption{10 Million Files}
\label{folder-read-throughputs-10M-files}
\end{subfigure}
\caption{Folder Read Throughput Distributed Over Time ($\mu$sec vs 20 buckets)} 

\end{figure*}

Each file’s write speed is captured, sorted, and saved in a bucket to determine the distribution pattern as shown in Figure 4. For example, majority of EXT4 file write speeds fall in two buckets: 10-15$\mu$s and 15-20$\mu$s whereas XFS write speed fall precisely 15$\mu$s regardless of number of files. Each of the file systems listed below requires a slightly different configuration to ensure it can accommodate one billion or more files and directories.

\textbf{EXT4} is the general-purpose filesystem for a majority of Linux distributions and gets installed as the default. However, it is unable to accommodate one billion files by default due to a small number of inodes available, typically set at 240 million files or folders, created during installation. To increase the inodes, the EXT4 filesystem needs to be reinstalled on a device with a custom configuration using the \textit{mkfs} command. The option -b 4096 creates block of size 4KB each resulting in about 1.2 billion inodes on a 14TB drive: \textit{mkfs.ext4 -N 1200005248 -b 4096 /dev/xxx}

\textbf{XFS} creates inodes dynamically as the files are being created but it requires that enough space is allocated to the inodes data structures on the disk, therefore XFS can handle one billion or more files and folders. This dynamic creation of inodes was observed when the progress bar momentarily pauses and then continues during the runs. For this purpose \textit{maxpct} flag is used with the command that allocates a percentage of the disk to the inode structure e.g. 10\% as shown in the command: \textit{mkfs.xfs -i maxpct=10 -f /dev/xxx}

\textbf{BtrFS} does not use inodes the like other filesystems therefore it always shows inode=0 for \textit{df} command. BtrFS has many other advanced features for a modern filesystem such as snapshots, fault tolerance, copy-on-write, support for huge size files, etc. The command to create BtrFS is: \textit{mkfs.btrfs -f /dev/xxx}

\textbf{ZFS} requires that a pool is created first and then it needs to be mounted. Only then it is available for write/read operations. A pool is created using: \textit{zpool create -f zfspoolname /dev/xxx}.

\textbf{F2FS} is a specialized filesystem to support SSDs. Upon installation, F2FS established a segmented disk layout. It uses inodes to points to files and folders. The command used was: \textit{mkfs.f2fs -i -s 10 -z 10 -f /dev/xxx}. This generated only 630 million inodes and unable to create 1 billion inodes so the test was restricted to 100 million files and corresponding folders.

The tables below captured performance statistics for writes and reads for the file systems listed above. The explanation of write performance numbers are as follows:

\begin{itemize}
  \item FWT\textsubscript{(min, ave, max)}: File Write Time\textsubscript{(min, ave, and max)} in $\mu$sec.
  \item WTh: Write Throughput in bytes per microsecond.
  \item TFWT: Total Folder Write Time is cumulative time to create (sub)folders in seconds.
  \item TfWT: Total File Write Time is time to create/write files in seconds.
  \item TWT: Total Write Time is the time to create folders, subfolders and files in seconds.
  \item FWs: Files written per second.
  \item BkWs: Blocks written per second. Typically 4KB for file systems but for ZFS it is 128KB.
  \item TFCWT: Total File Create for Write Time is the cumulative time to create files but does not include writing the data/bytes to the file.
  \item FCWT: Average time to create a files but not writing the bytes to the file.
  \item TByW: Total Bytes Written in gigabytes (GB).
  \item TBkW: Total 4KB or 128KB Blocks Written.
  \item DSU: Disk Space Used in gigabytes (GB).
  \item DSUO: Disk Space Utilization Overheads.
  \item Inodes: Inodes used to store folders, subfolders and files.
\end{itemize}

The explanation of read performance numbers are as follows:

\begin{itemize}
  \item FRT\textsubscript{(min, ave, max)}: File Read Time\textsubscript{(min, ave, and max)} in $\mu$sec.
  \item RTh: Read Throughput in bytes per microsecond.
  \item TFRT: Total Folder Read Time is cumulative time to search (sub)folders in seconds.
  \item TfWT: Total File Read Time is time to read files in seconds.
  \item TWT: Total Read Time is the time to search (sub)folders and files in seconds.
  \item FWs: Files read per second.
  \item BkWs: Blocks read per second. Typically 4KB for file systems but for ZFS it is 128KB.
  \item TFORT: Total File Open for Read Time is the cumulative time to open files but does not include reading the data/bytes from the file.
  \item FORT: Average time to open and read a files.
  \item TByR: Total Bytes Read in gigabytes (GB).
  \item TBkR: Total 4KB or 128KB Blocks Read.
  \item TRT: Total Run Time in seconds.
  \item CPUO: Computation Overheads is the total time taken by the run not including I/O.
  \item BkSize: Block size, typically 4KB except for ZFS's 128KB.
\end{itemize}

Total Runtim Time (TRT) can be loosely defined as Total Write Time (TFCWT + TWT), Total Read Time (TFORT + TRT) + processing overheads.

\section{Discussion}
The application generates folders, subfolders and files as per schedule in Table \ref{tab:read_write_schedule}. The files created with random data with normal distribution as shown in Figure \ref{fig:file-size-distribution} appended with a CRC-32 checksum computed using the random data generated for the file. Once all the files are generated, they are read back according to the same schedule. The file data is verified using CRC-32 read from the file and a generated CRC-32 checksum using the data from the file. Using EXT4 file system with 10 million files spread between 100 root folders and 1 subfolder each as the baseline to compare other file systems. The reason is to minimize the differences between each run and the configuration. All the  runs were carefully conducted. Linux Ubuntu 22.04 was out-of-the-box version with no configuration changes except regular system updates, re-creating the file system on the 14TB drive, rebooting the server between switching to different file system as well as in between runs when generating 1 billion files as these runs some times lasted 36 to 48 hours. Most of initial testing was conducted using EXT4 with 10 million files. Comparison analysis is performed between the baseline EXT4 and other file systems.

File write and read performance provided a glimpse of each file system and how they behave with different number of folders, subfolders, and files as each consumes an inode. As captured, the application is I/O-intensive, which means it spent majority of time waiting for disk I/O, i.e., writing or reading or searching the files. CPU overhead captured the numbers to show the percentage of the time system was *not* performing I/O. Starting with an initial assumption about worse write performance in comparison with read performance which turned out to be false because of Linux's built-in write caching mechanism. For example, EXT4 1B average file write time is only 14 $\mu$sec but the average read time is 62 $\mu$sec, about x3 times slower. ZFS 1B run is even worse in this regard since its average file write time is 16 $\mu$sec but the average file read time is 218 $\mu$sec, about x12 times slower. File read performance numbers provided a consistently interesting pattern as the graph shows. All small runs i.e., 10M files, for all filesystems, provided slower read speeds than write times. 

One of the questions that needed investigation was: Is there any performance degradation as the number of files in the filesystem increases? To answer this question, Creator code computes and uses folder performance over the entire run. Due to large number of folders it was difficult to show the trend over all folders, therefore it was decided to take 20 samples during the entire run spaced evenly.

The graphs in Figures \ref{folder-write-throughputs-all-files} and \ref{folder-read-throughputs-all-files} shows all filesystems with different runs. These graphs are further categorized by number of files generated or read to show different patterns at different times of the runs. The purpose is show the long term trends especially during the folders, subfolders and files creation. In some cases, write performance recovered and in some cases continue to deteriorate. Starting with EXT4, both read and write performance was consistent for all three runs though EXT4 shows momentary deterioration initially during 1 billion files run which was not reproducible so it was recorded as a possible anomaly. EXT4 gave relatively consistent write and read performances as well as long term performance degradation trends among the set of file system explored. 

XFS turned out to be about the same in performance and long term performance deterioration trend when compared with EXT4 (EXT4/XFS: WR:15/12, 24/28, 14/31 RD:10/5, 62/62, 62/73). In 1 billion folder read/write performance XFS was slower than EXT4 (see Figures \ref{folder-write-throughputs-1B-files} and \ref{folder-read-throughputs-1B-files}). 

We need to point out that there was an initial performance degradation in 100 million files run for both EXT4 and XFS which seems to be stabilized after create 25 folders and files. This degradation was not observed in 1 billion or 10 million files runs.

XFS turned out to be about the same in performance and long term performance deterioration trend when compared with EXT4 (EXT4/XFS: WR:15/12, 24/28, 14/31 RD:10/5, 62/62, 62/73).

In 1 billion folder read/write performance XFS was slower than EXT4 (see Figures \ref{folder-write-throughputs-1B-files} and \ref{folder-read-throughputs-1B-files}). We need to point out that there was an initial performance degradation in 100 million files run for both EXT4 and XFS which seems to be stabilized after create 25 folders and files. This degradation was not observed in 1 billion or 10 million files runs.

\begin{landscape}
\begin{table}[h]
\caption{File Write Performance Metrics}
\label{file-write-performance-metrics}
\centering
\begingroup
\setlength{\tabcolsep}{6pt} 
\renewcommand{\arraystretch}{1.2} 
\begin{threeparttable}
\begin{tabular}{lrrrrrrrrrrrrrr}
    \hlineB{2.7}
    \textbf{Filesystem} & \textbf{FWT\textsubscript{min,ave,max}} & \textbf{WTh} & \textbf{TFWT} & \textbf{TfWT} & 
    \textbf{TWT} & \textbf{FWs} & \textbf{BkWs} & \textbf{TFCWT} & \textbf{FCWT} & \textbf{TByW} & \textbf{TBkW} & \textbf{DSU} & \textbf{DSUO} & \textbf{Inodes} \\
    \hlineB{2.7}
    EXT4\textsubscript{10M\textsubscript{BL}}   & \small{5, 15, 2.53$\times 10^6$} & \small{348} & \small{0.02}   & \small{157}    & \small{157}    & \small{63} & \small{121}   & \small{253}    & \small{25} & \small{54.99}   & \small{19.18}   & \small{78.93}   & \small{30\%} & \small{10000200} \\
    EXT4\textsubscript{100M}                    & \small{5, 24, 2.50$\times 10^6$} & \small{224} & \small{0.12}   & \small{2452}   & \small{2452}   & \small{41} & \small{78}    & \small{3103}   & \small{31} & \small{549.95}  & \small{191.82}  & \small{789.29}  & \small{30\%} & \small{100001100} \\
    EXT4\textsubscript{1B}                      & \small{5, 14, 2.23$\times 10^6$} & \small{382} & \small{18.33}  & \small{14387}  & \small{14405}  & \small{70} & \small{133}   & \small{24762}  & \small{24} & \small{5507.59} & \small{1919.61} & \small{7898.75} & \small{30\%} & \small{1000010100} \\
    \hline
    XFS\textsubscript{10M}                      & \small{7, 12, 0.22$\times 10^6$} & \small{442} & \small{0.09}   & \small{124}    & \small{124}    & \small{80} & \small{154}   & \small{224}    & \small{22} & \small{54.99}   & \small{19.18}   & \small{138.34}  & \small{60\%} & \small{10000200} \\
    XFS\textsubscript{100M}                     & \small{7, 28, 0.76$\times 10^6$} & \small{191} & \small{0.31}   & \small{2871}   & \small{2872}   & \small{35} & \small{66}    & \small{2425}   & \small{24} & \small{549.96}  & \small{191.83}  & \small{903.31}  & \small{39\%} & \small{100001100} \\
    XFS\textsubscript{1B}                       & \small{7, 31, 0.72$\times 10^6$} & \small{176} & \small{8.07}   & \small{31208}  & \small{31216}  & \small{32} & \small{61}    & \small{26242}  & \small{26} & \small{5499.58} & \small{1918.25} & \small{8432.62} & \small{35\%} & \small{1000010100} \\
    \hline
    BtrFS\textsubscript{10M}                    & \small{6, 11, 0.14$\times 10^6$} & \small{464} & \small{0.01}   & \small{118}    & \small{118}    & \small{85} & \small{162}   & \small{220}    & \small{22} & \small{54.99}   & \small{19.18}   & \small{85.23}  & \small{35\%} & \small{0} \\
    BtrFS\textsubscript{100M}                   & \small{6, 12, 30.73$\times 10^6$} & \small{445} & \small{0.07}  & \small{1234}   & \small{1234}   & \small{81} & \small{155}   & \small{4968}   & \small{49} & \small{549.93}  & \small{191.82}  & \small{911.62} & \small{40\%} & \small{0} \\
    \hline
    F2FS\textsubscript{10M}                     & \small{4, 11, 0.19$\times 10^6$} & \small{495} & \small{0.01}  & \small{110}    & \small{110}   & \small{90} & \small{172}   & \small{1131}    & \small{113} & \small{55.00}  & \small{19.18}  & \small{118.99} & \small{54\%} & \small{10000400} \\
    F2FS\textsubscript{100M}                    & \small{4, 11, 0.39$\times 10^6$} & \small{471} & \small{0.05}  & \small{1166}   & \small{1166}  & \small{86} & \small{164}   & \small{11969}   & \small{119} & \small{549.97}  & \small{191.83}  & \small{1189.78} & \small{54\%} & \small{100003100} \\
    \hline
    ZFS\textsubscript{10M}                      & \small{11, 16, 0.05$\times 10^6$} & \small{337} & \small{0.02}  & \small{169}   & \small{163}   & \small{61} & \small{61}    & \small{394}   & \small{39} & \small{55.00}   & \small{10.00}   & \small{82.33}   & \small{33\%} & \small{10000200} \\
    ZFS\textsubscript{100M}                     & \small{11, 15, 0.09$\times 10^6$} & \small{358} & \small{5.92}  & \small{1535}  & \small{1541}  & \small{65} & \small{65}    & \small{3901}  & \small{39} & \small{549.94}  & \small{100.00}  & \small{823.22}  & \small{33\%} & \small{100001100} \\
    ZFS\textsubscript{1B}                       & \small{11, 16, 0.13$\times 10^6$} & \small{337} & \small{13.76} & \small{16316} & \small{16330} & \small{61} & \small{61}    & \small{40887} & \small{40} & \small{5499.59} & \small{1000.00} & \small{8232.76} & \small{33\%} & \small{1000010100} \\
    \hline
\end{tabular}
\begin{tablenotes}
All performance numbers are in microseconds. Table is divided in rows and columns: rows are for file system and number of files generated, whereas columns capture particular metrics. Each column is explained in detail the paper. EXT4 10M files is the baseline (EXT4\textsubscript{10M\textsubscript{BL}}). BtrFS file system does not use inodes therefore it is shown with 0 inodes. Unable to read back 1 billion files for BtrFS file system due to exponentially increasing read back times therefore it is not included in the write or read tables. Please note that folders are also represented by inodes.
\end{tablenotes}
\end{threeparttable}
\endgroup
\end{table}

\hfill 

\begin{table}[h]
\caption{File Read Performance Metrics}
\label{file-read-performance-metrics}
\centering
\begingroup
\setlength{\tabcolsep}{6pt} 
\renewcommand{\arraystretch}{1.2} 
\begin{threeparttable}
\begin{tabular}{lrrrrrrrrrrr|rrr}
    \hlineB{2.7}
    \textbf{Filesystem} & \textbf{FRT\textsubscript{min,ave,max}} & \textbf{RTh} & \textbf{TFRT} & \textbf{TfRT} & \textbf{TRT} & \textbf{FRs} & \textbf{BkRs} & 
    \textbf{TFORT} & \textbf{FORT} & \textbf{TByR} & \textbf{TBkR} & \textbf{TRT} & \textbf{CPUO} & \textbf{BkSize} \\
    \hlineB{2.7}
    EXT4\textsubscript{10M\textsubscript{BL}}   & \small{2, 10, 4.12$\times 10^6$}  & \small{505}  & \small{0.00}   & \small{108}   & \small{108}   & \small{92}  & \small{176} & \small{23}    & \small{2}  & \small{54.99}     & \small{19.18}   & \small{683.66}     & \small{21\%} & \small{4096} \\
    EXT4\textsubscript{100M}                    & \small{2, 62, 1.73$\times 10^6$}  & \small{88}   & \small{0.05}   & \small{6226}  & \small{6226}  & \small{16}  & \small{30}  & \small{1688}  & \small{16} & \small{549.95}    & \small{191.82}  & \small{15049.99}   & \small{10\%} & \small{4096} \\
    EXT4\textsubscript{1B}                      & \small{2, 62, 0.95$\times 10^6$}  & \small{88}   & \small{185.42} & \small{62200} & \small{62386} & \small{16}  & \small{30}  & \small{20181} & \small{20} & \small{5507.59}   & \small{1919.61} & \small{136914.58}  & \small{11\%} & \small{4096} \\
    \hline
    XFS\textsubscript{10M}                      & \small{2, 5, 2.16$\times 10^6$}   & \small{1028} & \small{0.00}   & \small{53}    & \small{53}    & \small{187} & \small{358} & \small{23}    & \small{2}  & \small{54.99}     & \small{19.18}   & \small{571.45}     & \small{26\%} & \small{4096} \\
    XFS\textsubscript{100M}                     & \small{3, 62, 1.17$\times 10^6$}  & \small{87}   & \small{0.09}   & \small{6285}  & \small{6285}  & \small{16}  & \small{30}  & \small{1351}  & \small{13} & \small{549.96}    & \small{191.83}  & \small{14560.15}   & \small{11\%} & \small{4096} \\
    XFS\textsubscript{1B}                       & \small{37, 73, 1.17$\times 10^6$} & \small{74}   & \small{253.65} & \small{73852} & \small{74105} & \small{14}  & \small{25}  & \small{39588} & \small{39} & \small{5499.58}   & \small{1918.25} & \small{188036.72}  & \small{9\%}  & \small{4096} \\
    \hline
    BtrFS\textsubscript{10M}                    & \small{2, 4, 0.0$\times 10^6$}    & \small{1349} & \small{0.00}   & \small{40}    & \small{40}    & \small{245} & \small{470} & \small{24}    & \small{2}  & \small{54.99}     & \small{19.18}   & \small{553.70}     & \small{27\%} & \small{4096} \\
    BtrFS\textsubscript{100M}                   & \small{2, 70, 0.62$\times 10^6$}  & \small{77}   & \small{0.08}   & \small{7071}  & \small{7071}  & \small{14}  & \small{27}  & \small{1759}  & \small{17} & \small{549.93}    & \small{191.82}  & \small{16674.51}   & \small{10\%} & \small{4096} \\
    \hline
    F2FS\textsubscript{10M}                     & \small{1, 2, 0.01$\times 10^6$}    & \small{2038} & \small{0.00}  & \small{26}    & \small{26}    & \small{371} & \small{710} & \small{23}    & \small{2}   & \small{55.00}    & \small{19.18}   & \small{1436.08}    & \small{10\%} & \small{4096} \\
    F2FS\textsubscript{100M}                    & \small{2, 131, 0.66$\times 10^6$}  & \small{41}   & \small{1.36}  & \small{13163} & \small{13164} & \small{8}   & \small{14}  & \small{20780} & \small{207} & \small{549.97}   & \small{191.83}  & \small{48771.11}   & \small{3\%}  & \small{4096} \\
    \hline
    ZFS\textsubscript{10M}                      & \small{11, 16, 0.06$\times 10^6$}  & \small{332} & \small{0.00}   & \small{165}    & \small{165}    & \small{61} & \small{60} & \small{30}    & \small{3}  & \small{55.00}    & \small{10.00}   & \small{912.24}     & \small{17\%} & \small{131072} \\
    ZFS\textsubscript{100M}                     & \small{83, 217, 0.65$\times 10^6$} & \small{25}  & \small{2.91}   & \small{21707}  & \small{21710}  & \small{5}  & \small{4}  & \small{3961}  & \small{39} & \small{549.94}   & \small{100.00}  & \small{32981.95}   & \small{6\%}  & \small{131072} \\
    ZFS\textsubscript{1B}                       & \small{73, 230, 0.83$\times 10^6$} & \small{23}  & \small{10.52}  & \small{230447} & \small{230457} & \small{4}  & \small{4}  & \small{41671} & \small{41} & \small{5499.59}  & \small{1000.00} & \small{348312.94}  & \small{5\%}  & \small{131072} \\
    \hline
\end{tabular}
\begin{tablenotes}
All performance numbers are in microseconds. Table is divided in rows and columns: rows are for file system and number of files read, whereas columns captures a particular metrics. Each column is explained in detail the paper. Right most three columns are included though they are common for write and read operation. Total Run Time (TRT) is the total run time of the application against a target file system. CPUO is the CPU bound processing overheads that excludes any I/O. BlkSize is the block size used by the file system.
\end{tablenotes}
\end{threeparttable}
\endgroup
\end{table}

\end{landscape}

To capture the long term trend to assess performance deterioration, we captured write and read performance of every 5th folder created and read back and stored these numbers in 20 buckets. 
BtrFS testing took more time than all others file system testing combined because its read performance deteriorated to a point that 1 billion files run never got completed even after running for 72 hours straight.
 
The write portion of the runs completed but reads were too slow to over 60 minutes and gradually increasing just to read one folder so runs were killed, rebooted, re-installed file system and still failed. BtrFS file write was comparable to EXT4 but the file read performance was lightly lower (around 65 $\mu$sec). We did not observe any long term write or read performance degradation for BtrFS for 100 million and 10 million files runs.

F2FS, though meant for flash drives, was tested on 14TB HDD to be consistent with other file systems under test. It was not expected we could create 1 billion files but we tested F2FS anyway by increasing the inodes. We were unable to increase beyond 600 million so restricted F2FS to 100 million files maximum. F2FS was slightly better than EXT4 in file write performance but in file read performance F2FS was slower than EXT4 in 100 million files runs. In folder write and read long term trend, F2F2 was stable in both 100 million and 10 million files runs.

ZFS was different in terms of its installation. Once installed it behaved like any other system, i.e., no changes to the application code required. ZFS file write performance was very consistent (16/15/16 $\mu$sec) but file read performance was much higher (16/217/230 $\mu$sec). ZFS runs took much higher clock time than other file systems. For 1 billion files write and read back it took 330 thousand seconds, about 92 hours, to complete.

CPU overheads are defined as time the code spend not performing any I/O, such as calculating statistics, or other housekeeping work. The decision to switch to pure C from Java was deliberate in order to avoid any overhead imposed by JVM or Java programming constructs. The number captured showed a consistent pattern: ~20\% for 10 million files runs, ~10\% for 100 million files runs, and ~10\% for 1 billion files runs. CPU overheads were computed using Total Run Time (TRT), Total Write Time (TWT), Total File Create for Write Time (TFCWT), Total Read Time (TRT), and Total File Open for Read Time (TFORT). $\textrm{CPUO = TRT - (TWT + TFCWT + TRT + TFORT) / TRT}$

\section{Conclusion}

This study investigated and compared several popular Linux filesystems: EXT4, XFS, BtrFS, F2FS, and ZFS for their ability to create and manage one billion files and measure file creation and reading timing and any performance degradation during and after creating the files. Except for F2FS, all filesystems were able to handle one billion files after increasing the inodes and re-mounting the filesystem i.e., no other operating system onfiguration changes were made. BtrFS was tested with 1 billion files but reading these files were too slow to capture the final performance metrics. To exercise these file systems and capture desired metrics, a tool is developed (initially in Java) in C programming language. The second decision was made to use an on-prem server rather than using the cloud to avoid unnecessary round-trip delays.

The decision to pick popular Linux filesystems was based on a couple of factors: their availability under most Linux distributions, non-clustered, and at least one, F2FS, supports SSDs natively. EXT4 comes with many Linux distributions by default therefore it was chosen to establish the baseline for performance benchmarking. XFS, BtrFS and ZFS were chosen because they were comparable in features with EXT4 for this paper.

All three filesystems, except F2FS, were able to handle one billion files. Once they ran successfully, and essential metrics was captured, these large tests were not repeated because they took days to complete. One of the fundamental limitations that were discovered was the default number of inodes. EXT4 needed to be configured and reinstalled whereas XFS, BtrFS and ZFS handle inode creation automatically. However, XFS requires that when installing the filesystem, a certain percentage of storage must be dedicated to inode data structures. EXT4 is a good choice if a large number of small files need to be stored on the filesystem. XFS has better block-based performance if performance is the primary requirement.

Original questions that were asked in the beginning are partially answered--ability to create one billion files on selected filesystems except for F2FS due to the lack of needed inodes. In addition, the study was able to answer about performance degradation. Except for EXT4, other filesystems show some signs of performance deterioration as the number of files increased, BtrFS was the worse in this regard. In addition, disk utilization overhead were observed for all four filesystems, XFS has the worse overheads followed by F2FS, useful for storage estimation by system designers.

The study captured extensive amounts of data for all four filesystems including individual read and write speeds as well as cumulative reads and writes performance, read, and write throughput, the number of files written and read per second, disk block utilization, and disk utilization overheads. Further work can be conducted to measure performance using real-world databases on these filesystems with high data volume and high transaction rates. Additional statistical analysis can be performed by working with the raw data captured to establish a statistical basis for the findings using techniques to include Confidence Intervals, Paired Observations, Linear Regression, Hypothesis Testing, and Analysis of Variance (ANOVA). We hope that study is useful for system designers in selecting the right file system for thier purpose.

\section*{Acknowledgment}
I would like to acknowledge my professor Dr. Yue Cheng (mrz7dp@virginia.edu) who provided guidance and timely advice to conduct this study and in the preparation of this paper.


\end{document}